Beryllium nitride thin film grown by reactive laser ablation

G. Soto, J. A. Díaz, R. Machorro, A. Reyes-Serrato*
Centro de Ciencias de la Materia Condensada, UNAM, A. Postal 2681, 22800 Ensenada B.C., México.

W. de la Cruz
Centro de Investigación Científica y de Educación Superior de Ensenada, Km. 107 carretera Tijuana-Ensenada, 22800 Ensenada B.C., México.

Beryllium nitride thin films were grown on silicon substrates by laser ablating a beryllium foil in molecular nitrogen ambient. The composition and chemical state were determined with Auger (AES), X-Ray photoelectron (XPS) and energy loss (EELS) spectroscopies. A low absorption coefficient in the visible region, and an optical bandgap of 3.8 eV, determined by reflectance ellipsometry, were obtained for films grown at nitrogen pressures higher than 25 mTorr. The results show that the reaction of beryllium with nitrogen is very effective using this preparation method, producing high quality films.

*keywords:* Beryllium nitride, laser ablation, thin films, wide bandgap, optical properties.

PACS numbers: 81.05.Je, 81.15.Fg, 81.40.Tv

*Corresponding author. Centro de Ciencias de la Materia Condensada UNAM, P.O. Box 439036, San Ysidro CA 92143. Tel. ++52-6-174 4602. Fax. ++52-6-174 4603. E-mail address: armando@ccmc.unam.mx

The production and characterization of nitrides materials are subject of many publications due to technological importance. An important example is the research done to obtain a material suitable for laser diodes emitting in the blue/ultraviolet region. A very important feature is the quantum efficiency, therefore the material must own a direct band-gap. In a recent publication it is reported the theoretical direct bandgap of beryllium nitride alpha phase, $\alpha$-$Be_3N_2$, which is expected to be in the range of 4.05-4.47 eV [1].

There are few reports on the preparation of bulk $Be_3N_2$ [2]. Nevertheless, a previous work has demonstrated that laser-induced thin films of beryllium nitride is possible at the beryllium/liquid nitrogen interface [3]. In this work, we demonstrate that nitrogen can be incorporated into beryllium films by ablating a beryllium target in a background $N_2$ gas atmosphere. This method has been successfully used for growing high-density films, as titanium nitride and silicon nitride [4].

With this aim, we have prepared a series of films by ablating a high purity beryllium foil at different pressures of $N_2$. The experimental set up has been described in detail in a previous paper [5]. Essentially, it is consist of a UHV chamber, an excimer laser (KrF, $\lambda$ = 248 nm), real time ellipsometry and several *in situ* electronic spectroscopies (AES, XPS, ELS) in a Riber LDM-32 system. Laser energy, number of pulses and pulse repetition rate were kept fixed at *400 mJ, 2000* pulses and *2 Hz*, respectively.

The auger spectrum of the film deposited at nitrogen pressures ($P_N$) of 13 mTorr, Fig. 1, shows that the only signals proceed from beryllium, nitrogen and oxygen. The occurrence of oxygen can be attributed to residual water in the nitrogen introduction lines. This analysis confirm that nitrogen incorporation into the films is effective, and the *Be* KVV transition is comparable of those reported by G. Hanke and K. Müller for $Be_3N_2$ [6].

The EELS spectra for films grown at $P_N$ of 0, 1, 13 and 50 mTorr are presented in Fig. 2. For films growth at $P_N$ = 0 mTorr, two main peculiarities are found: the maximum peak intensity at 19.2 eV, corresponding to free electron bulk plasmon oscillations of metallic beryllium; and the surface plasmon, observed as a shoulder of the main peak in the 12-14 eV region. As the nitrogen pressure is increased, the bulk plasmon shifts from 19.2 eV ($P_N$ = 0 mTorr) up to 23 eV ($P_N \geq$ 13 mTorr). This phenomena is similar to the formation of silicon nitride, where the silicon 'metallic' bulk plasmon is found at 17 eV, and after $Si_3N_4$ formation, it moves up to 23 eV [7]. This is due to higher density of electronic states in the valence band for the nitride phase in relation to the metallic one.

High-resolution XPS spectra around the $Be_{1s}$ core level are shown in Fig. 3 for films grown at different $P_N$. The binding energy ($E_b$) for film grown at $P_N$ = 0 mTorr was found at 111.8 eV, which is associated to the beryllium in metallic state [8]. As the nitrogen is introduced in the chamber, the intensity of this peak decreases and another peak is observable at higher $E_b$. At $P_N$ = 5 mTorr, the spectrum is clearly composed of two main peaks, it is indicative of the formation of a new phase with a singular chemical state. At higher nitrogen pressures, the peak corresponding to high binding energy becomes predominant. For pressures beyond of 25 mTorr, there are no significant changes in the energy and peak intensity. This new $Be_{1s}$ state has a binding energy of 114.0 eV, which could be assigned to a new ionic phase, where the beryllium has transferred his valence electrons to the electron-acceptor nitrogen atoms. In beryllium oxide the binding energy, determined by XPS [8] is 113.7, which is, within the instrument resolution, approximately the same energy that we found in our experiment. Subsequently, the total electron transfer from *Be* atoms is identical in beryllium nitride than in beryllium oxide. In addition, the maximum intensity in the $N_{1s}$ XPS spectrum was measured at 397.4 eV, consequently, the ionicity degree in this compound is considerably high.

In order to study the modification of stoichiometry into the film as a function of $N_2$ pressure, the XPS peaks area was obtained after linear background subtraction for $Be_{1s}$, $N_{1s}$ and $O_{1s}$ core levels. In Fig. 4, can be observed that nitrogen and beryllium signals in the films steadily change from $1.3$ mTorr $\leq P_N \leq 25$ mTorr. These plots reach a plateau for $P_N$ higher than $25$ mTorr. It is worth to note that when one atomic component change from a pure metallic to an ionic state, the classical semi-quantitative XPS approach is invalid since the tabulated sensitivity factors are considerably altered. To make use of the XPS peak intensity to determine the atomic concentration, a rigorist and genuine quantitative approach must be applied. With this aim, we presuppose that in the saturation region the produced material is homogeneous and effectively $Be_3N_2$. Then, the XPS intensity from a core level *k* can be written as:

$$I_k \ \alpha \ I_0 n \sigma_k \lambda_{MED}(E_k) T(E_k), \qquad (1)$$

where $n$ is the atomic density in the sample, $\sigma_K$ is the photoionization cross-section for level $k$ [9], $\lambda_{MED}(E_k)$ is the mean escape depth for electron with kinetic energy $E_K$ in the examined material [10], and $T(E_K)$ is the spectrometer transmission function [11]. $I_0$ is a constant factor, which depends on the x-ray radiation intensity. The product $\sigma_k \lambda_{MED}(E_k)T(E_k)$ is named as the theoretical relative sensitivity factor, $S_r$, for the core level $k$. All these quantities were calculated for $Be_{1s}$ and $N_{1s}$ in beryllium nitride using the *e*-analyzer CAMECA Mac-3 transmission function with an $Al\ K_\alpha$ x-ray source. The ratio calculated was $S_r^N / S_r^{Be} = 9.36$. With this evaluation, film stoichiometry in the saturation region was calculated, yielding a result of 60.6% and 39.4% (± 2%) for beryllium and nitrogen concentration respectively, in a good agreement with the anticipated presumption.

Optical properties of the films were determined *in situ* by reflectance ellipsometry. The real and imaginary parts ($n$, $k$) of the refractive index, Fig. 5, are derived from numerical fitting to the experimental data ($\psi$,$\Delta$) of the vacuum/film/substrate system as a function of photon energy [12]. The film grown at $P_N$ = 25 mTorr presents refractive index in the range 2.09 at 630nm to 2.38 at 275nm. The imaginary part is negligible in the visible region, 750 to 400nm, and rapidly increases below that wavelength. The absorption coefficient, $\alpha$, is related to the extinction coefficient, $k$, by $\alpha = (4\pi / hc)\ (E\ k)$, where $h$ is the Planck constant, $c$ is the speed of light, and $E$ is the photon energy. The coefficient $\alpha$ follow the Tauc's behavior for amorphous materials, where the optical band gap, $E_g$, can be obtained by intersecting the straight line behavior at the high absorption region, $(\alpha E)^{1/2}$ , with the *E*-Axis [13], as it is shown in the inset of Fig. 5. The values of $E_g$ for the film grown at 25 mTorr of $N_2$ was determined at *3.8* eV.

In summary, beryllium nitride thin films have been grown by ablating a beryllium foil in a $N_2$ atmosphere at different pressures. Micrographs acquired by scanning electron microscopy, not shown here, reveal films morphologically smooth and homogeneous, except by a few droplets, which are typical in films processed by laser ablation. Good adherence of the films to silicon substrates was also noticed. We could modulate the stoichiometry of the films varying the nitrogen pressure in the 1 mTorr $\leq P_N \leq$ 25 mTorr range. At higher pressures, the stoichiometry remains unchanged. In this region, the film composition can be satisfactorily given by $Be_3N_2$. The difference of 2.2 eV in $\Delta E_B$ of $Be_{1s}$, and the increment in the plasmon energy are evidences that beryllium nitride has been successfully made. In a previous work, which suggest that $\alpha$-$Be_3N_2$ has a direct gap, and our optical measurement of $E_g$ = 3.8 eV band gap, acquire importance due to their potential applications in optoelectronics.

The authors are grateful to Israel Gradilla and Margot Sainz for technical assistance. W. de la Cruz acknowledges the scholarship from COLCIENCIAS (Colombia).

Figure captions.

FIG. 1.- AES spectrum for a film grown by ablating a beryllium target at 13 mTorr of $N_2$.

FIG. 2.- EELS spectra for films growth at $P_N$ of (a) 0 mTorr; (b) 1 mTorr; (c) 13 mTorr; and (d) 50 mTorr.

FIG. 3.- XPS spectra in the $Be_{1s}$ region for films growth at 0 mTorr $\leq P_N \leq$ 100 mTorr.

FIG. 4.- Atomic concentration determined by XPS as a function of the nitrogen gas pressure.

FIG. 5.- Real and imaginary parts ($n$,$k$) of the refractive index *vs* photon energy of a film of beryllium nitride grown at $P_N$ = 25 mTorr. (Inset) Tauc's plot showing the experimental determination of the optical bandgap of the same film.

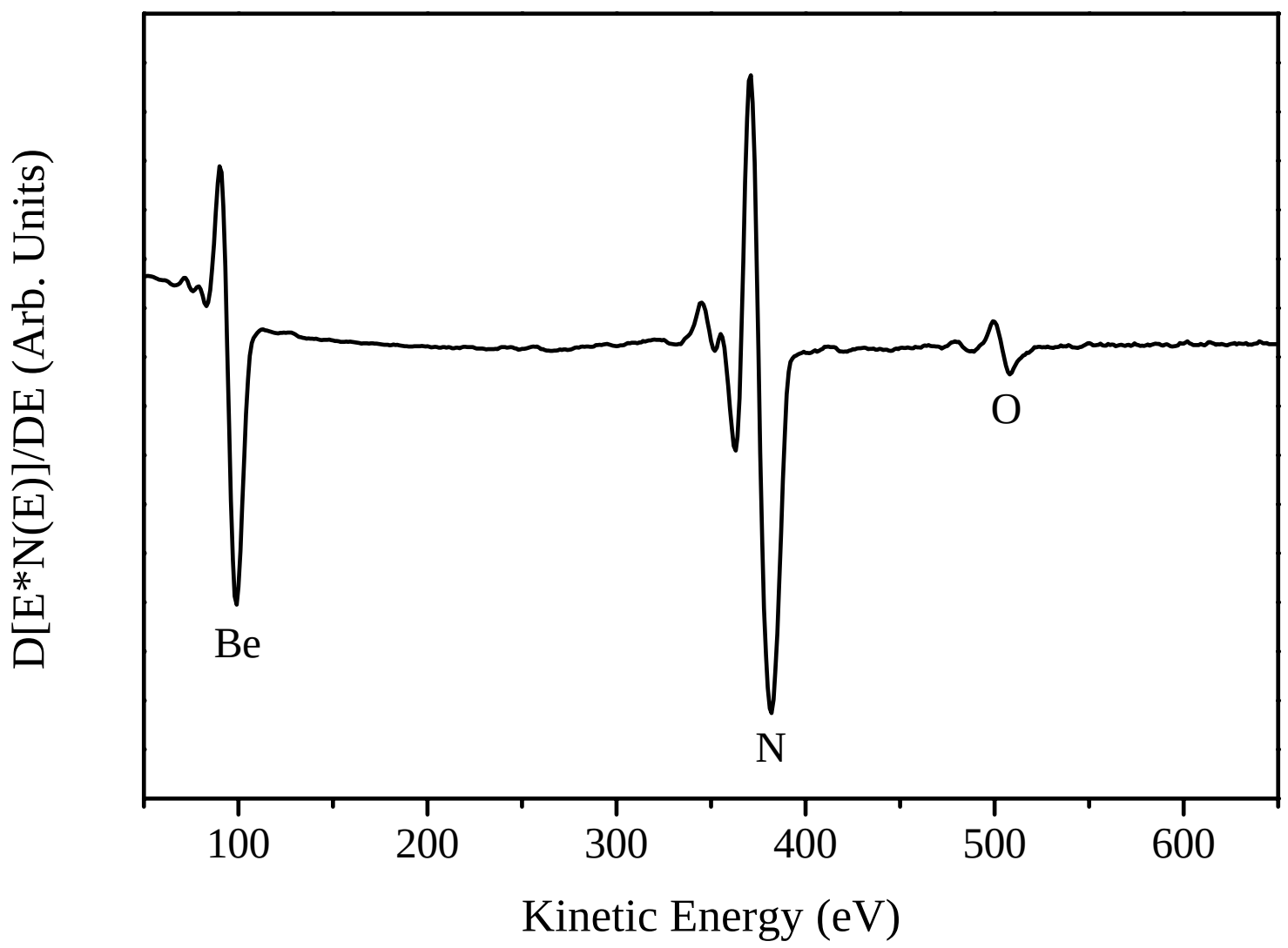

Fig. 1

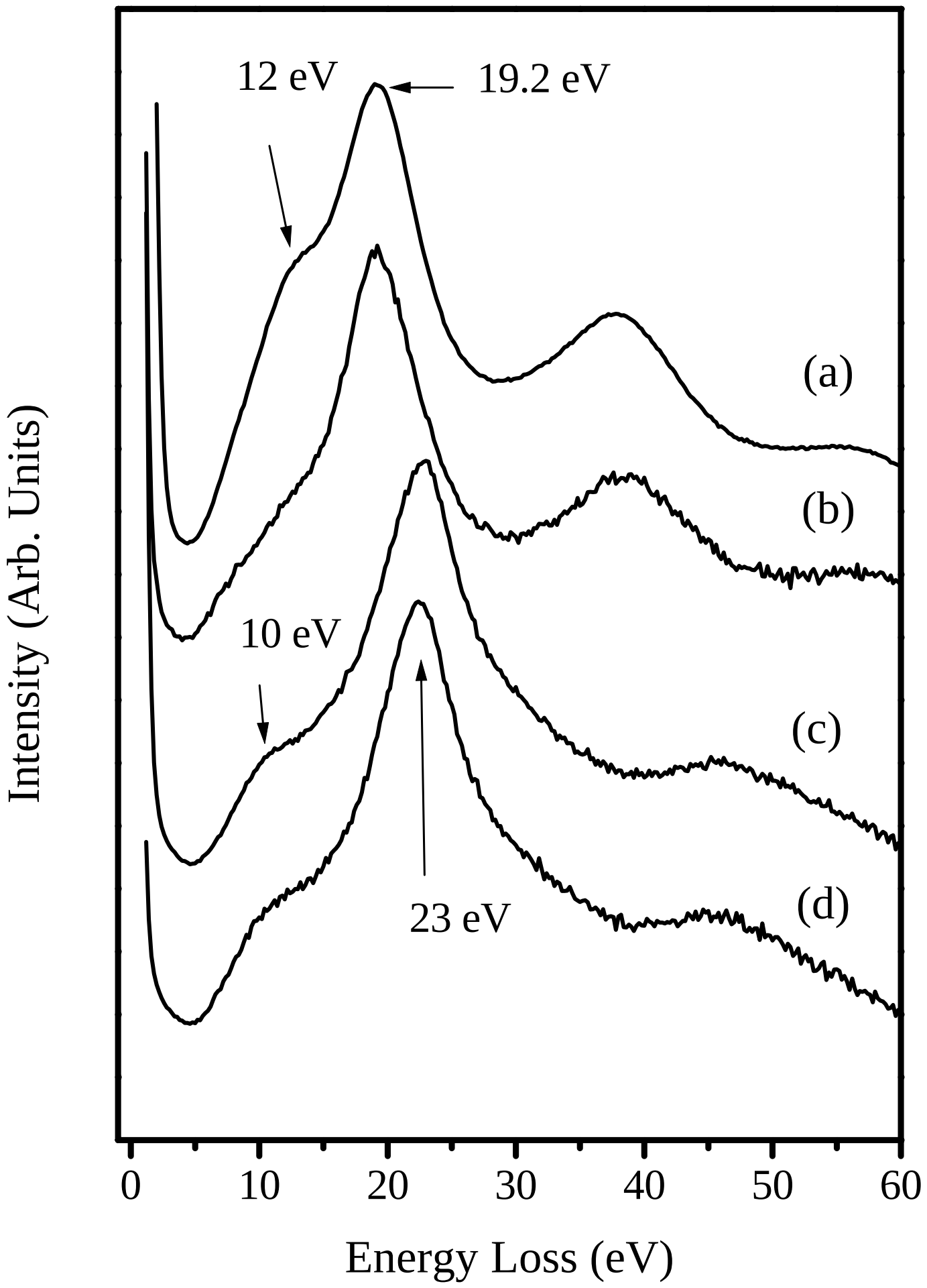

Fig. 2

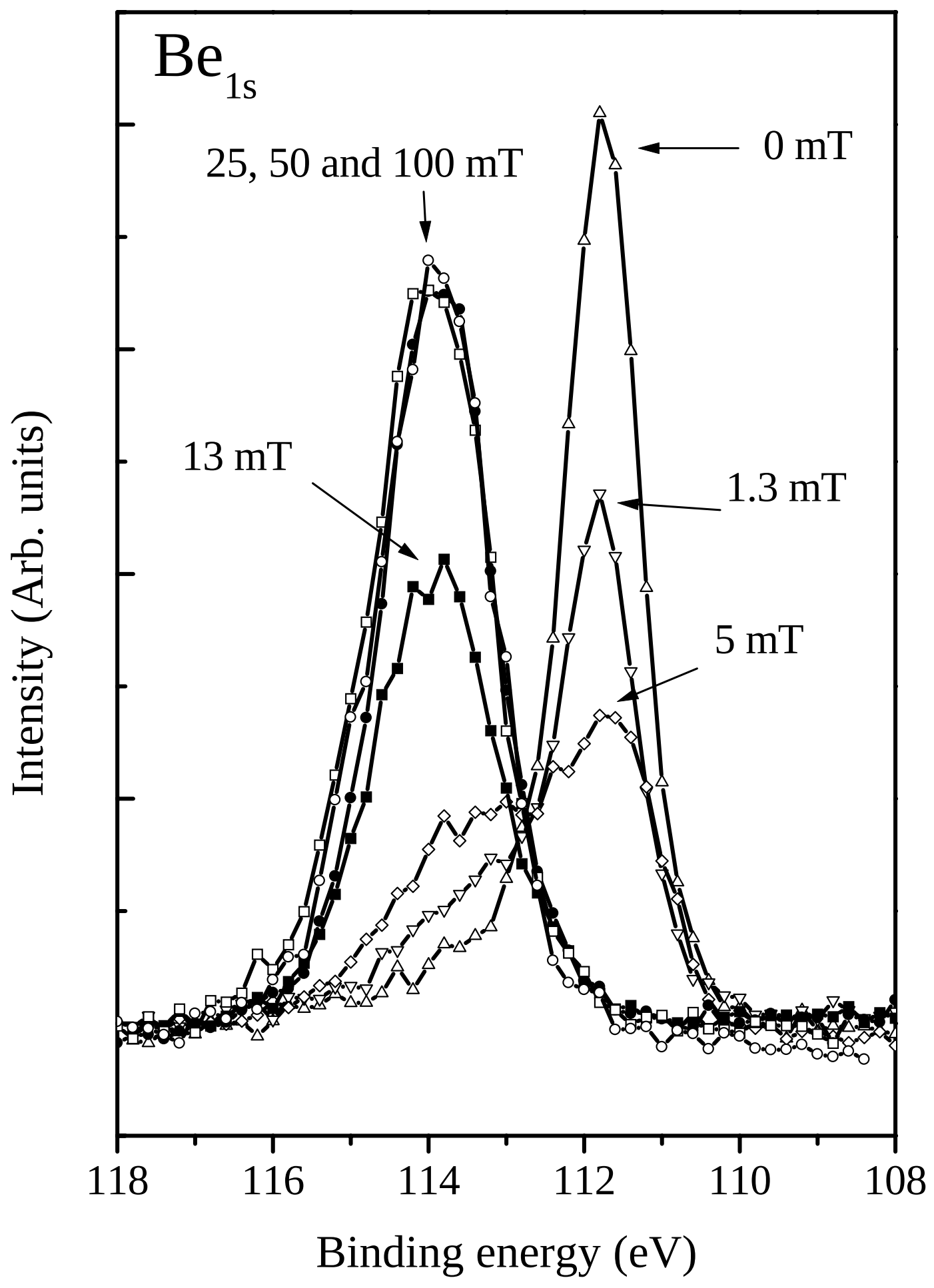

Fig. 3

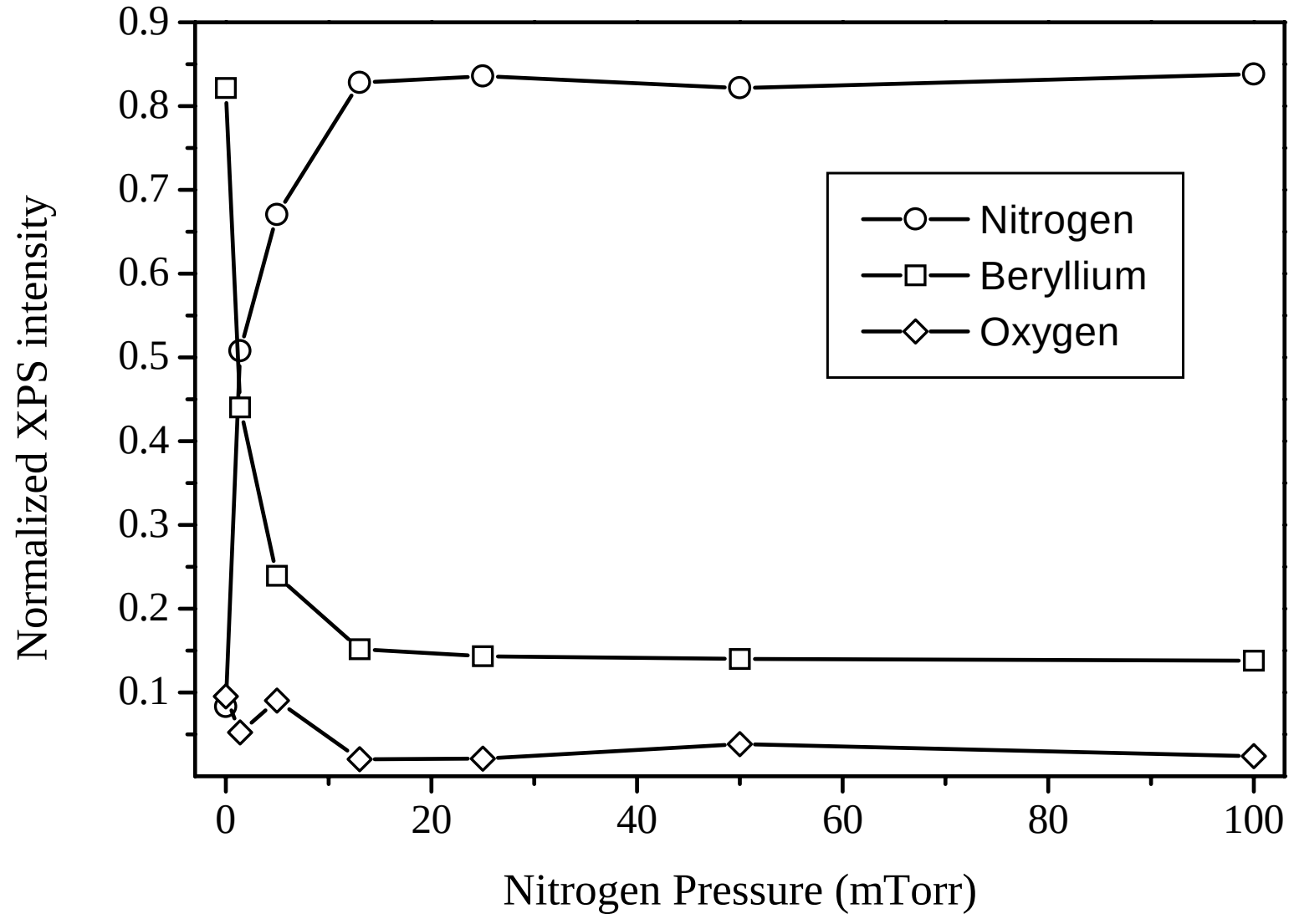

Fig. 4

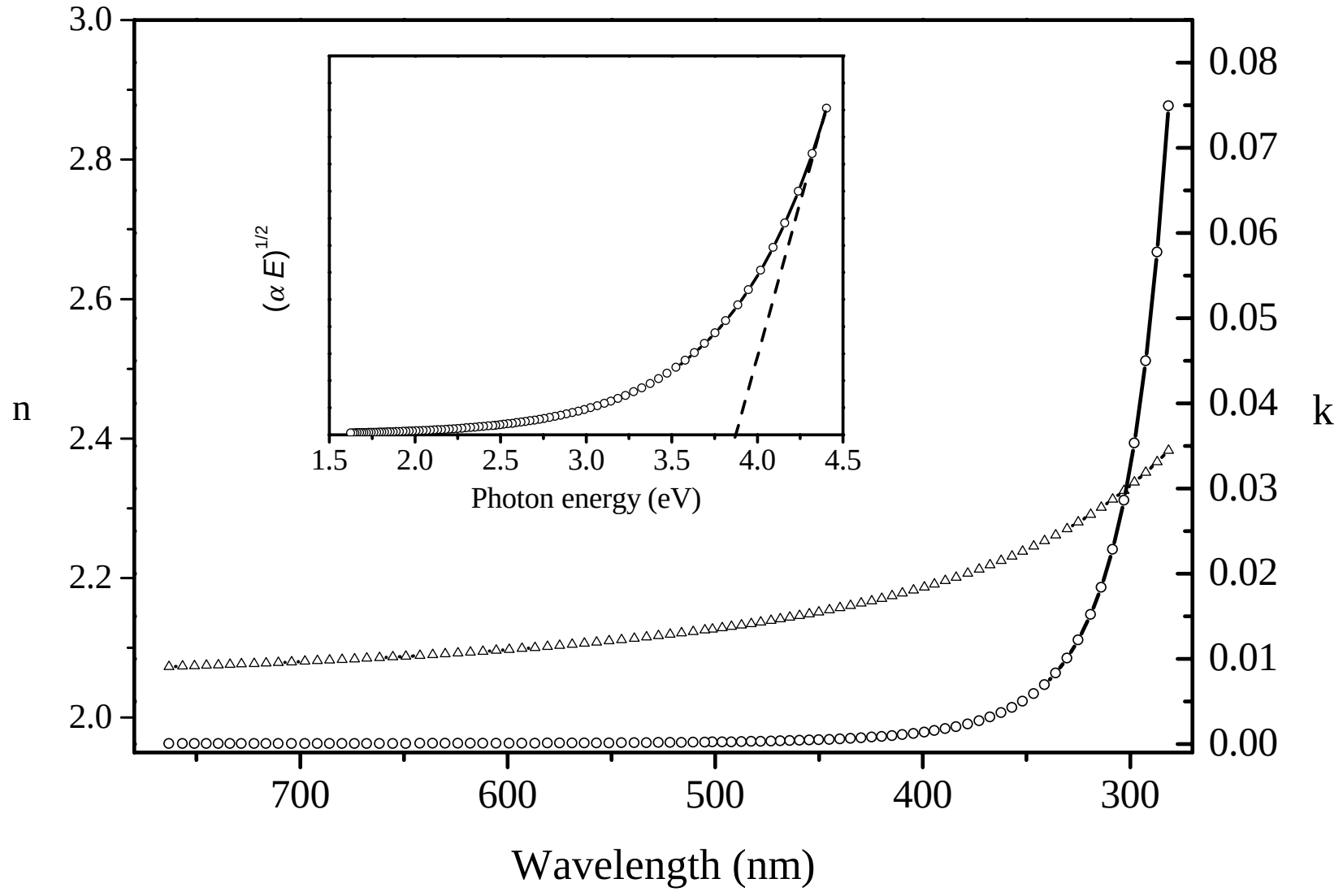

Fig. 5